\documentclass[prl,superscriptaddress,twocolumn,showpacs,preprintnumbers,amsmath,amssymb]{revtex4}

\usepackage{graphicx}% Include figure files
\usepackage{dcolumn}% Align table columns on decimal point
\usepackage{bm}% bold math
\usepackage{eucal}
\usepackage{ulem}

\begin{document}

\title{Phonon-induced quadrupolar
ordering of the magnetic superconductor TmNi$_2$B$_2$C}
\author{N.~H.~Andersen}
\affiliation{Materials Research Department, Ris\o{} National
Laboratory, DK-4000 Roskilde, Denmark}
\author{J.~Jensen}
\affiliation{Niels Bohr Institute, Universitetsparken 5, DK-2100
Copenhagen \O, Denmark}
\author{T.~B.~S.~Jensen}
\affiliation{Materials Research Department, Ris\o{} National
Laboratory, DK-4000 Roskilde, Denmark}
\author{M.~v.~Zimmermann}
\affiliation{Hamburger Synchrotronstrahlungslabor HASYLAB at
Deutsches Elektronen Synchrotron DESY, Notkestrasse 85, 22603
Hamburg, Germany}
\author{R.~Pinholt}
\author{A.~B.~Abrahamsen}
\author{K.~N\o rgaard~Toft}
\affiliation{Materials Research Department, Ris\o{} National
Laboratory, DK-4000 Roskilde, Denmark}
\author{P.~Hedeg\aa rd}
\affiliation{Niels Bohr Institute, Universitetsparken 5, DK-2100
Copenhagen \O, Denmark}
\author{P.~C.~Canfield}
\affiliation{Ames Laboratory and Department of Physics and Astronomy,
Iowa State University, Ames, Iowa 50011, USA}

\date{\today}

\begin{abstract}
We present synchrotron x-ray diffraction studies revealing that the
lattice of thulium borocarbide is distorted below $T_Q^{}\simeq13.5$
K at zero field. $T_Q^{}$ increases and the amplitude of the
displacements is drastically enhanced, by a factor of 10 at 60 kOe,
when a magnetic field is applied along [100]. The distortion occurs
at the same wave vector as the antiferromagnetic ordering induced by
the $a$-axis field.  A model is presented that accounts for the
properties of the quadrupolar phase and explains the peculiar
behavior of the antiferromagnetic ordering previously observed in
this compound.
\end{abstract}

\maketitle

More than ten years ago it was discovered that four of the rare-earth
borocarbides (Tm, Ho, Er, Dy) shows coexistence of superconductivity
and antiferromagnetic ordering with comparable transition
temperatures \cite{nagarajan94,cava94}. Superconductivity occurs
below $T_c^{}=11$ K and the N\'eel temperature is
$T_{\text{N}}^{}=1.52$ K in TmNi$_2$B$_2$C \cite{Movs}. The
characterization and the understanding of these intermetallic
compounds have made much progress during the passed decade, see the
recent review \cite{Rev05}. Although unusual phenomena have been
detected, the type-II superconductivity of these materials seems to
be described by the BCS-theory. The electronic system strongly
influences the acoustic and optical $\Delta_4^{}$-phonon branches
close to $(0.5,0,0)$ and $(0,0.5,0)$ in the non-magnetic Lu and Y
versions of these compounds \cite{dervenagas95,kawano96,bullock98}.
The two vectors connect parallel areas of the Fermi surface
\cite{dugdale99}, so nesting explains why the electron--phonon
interaction is particularly large at these wave vectors. This
enhancement of the electron--phonon interaction is probably the
primary reason for the relatively high $T_c^{}$-values of the
borocarbides. Nesting is generally also assumed to be important for
the magnetic susceptibility of the band electrons by causing a
maximum in the Ruderman--Kittel--Kasuya--Yoshida (RKKY) coupling of
the rare-earth moments at the nesting vectors \cite{dugdale99}.

The magnetic properties of TmNi$_2$B$_2$C single crystals have been
studied by magnetization measurements \cite{cho95} and by neutron
diffraction \cite{Sternlieb,Katrine00,Katrine04}. The ordering vector
at zero field is $\bm{Q}_F^{}=(0.094,0.094,0)$ with the magnetic
moments transversely polarized along the easy $c$ axis. The shift of
the magnetic ordering vector away from zero is probably a consequence
of the Anderson-Suhl screening of the long-wavelength susceptibility
of the superconducting electrons \cite{Katrine00}. Applying a field
in excess of 10-15 kOe along $[100]$, the magnetic ordering changes
into another antiferromagnetic one at the wave vector
$\bm{Q}_A^{}=(0.483,0,0)$. The new magnetic phase was considered to
be the stable one at zero field in case the electrons were normal,
but because the superconducting electrons were more strongly affected
by the superzone energy gaps at $\bm{Q}_A^{}$ than at $\bm{Q}_F^{}$,
the long-wavelength ordering would become the most favorable one of
the combined system at small fields. In the original experiment
\cite{Katrine00} only fields of up to 18 kOe were applied. The
extension of the experiment up to a field of 60 kOe \cite{Katrine04}
clearly showed that this conjecture could not be correct, since the
$\bm{Q}_A^{}$-phase becomes the more stable the higher the field is,
and the temperature variation more and more resembles that of the
antiferromagnetic order being induced by the uniform field. This
highly surprising observation motivated our search for a
field-dependent quadrupolar ordering of the system, observable
through the accompanying deformation of the lattice.

\begin{figure*}[t]
\includegraphics[width=0.99\linewidth]{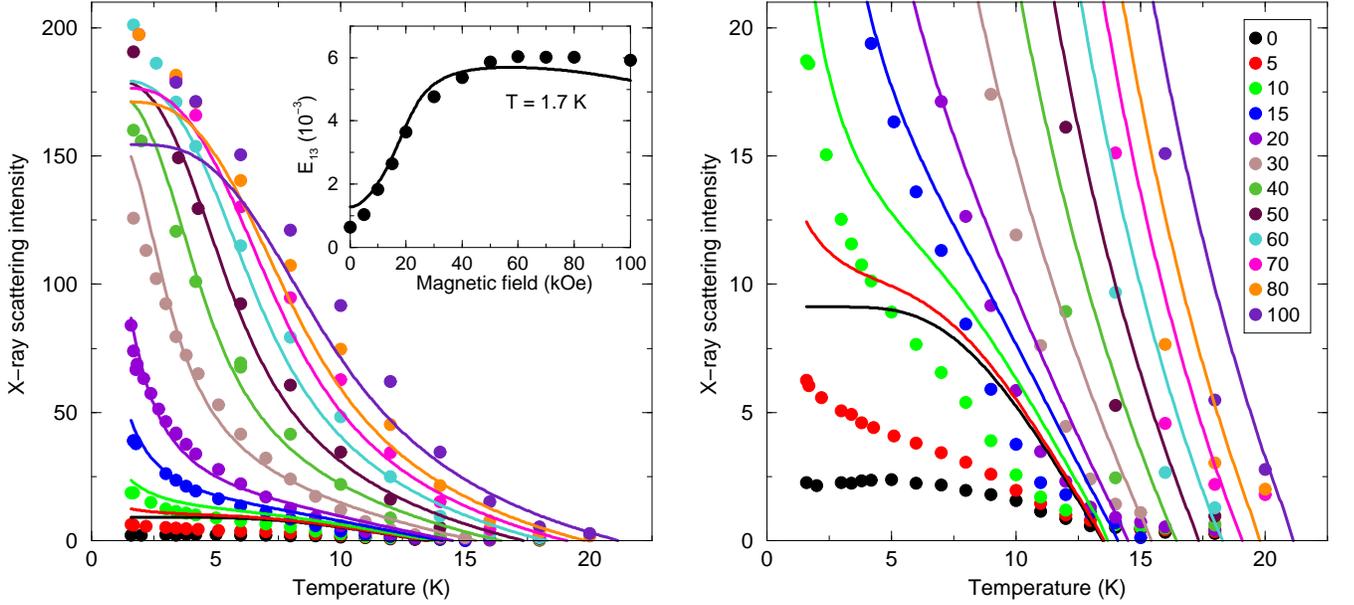}
\caption{The integrated x-ray scattering intensity at $(0.484,0,8)$
as a function of temperature in the presence of an applied field
along $[100]$. The colored symbols used for the different values of
the field (in units of kOe) are specified in the legend box. The two
figures include the same results, but the intensity scales differ by
a factor of 10. The intensity is proportional to the square of the
displacement of the Tm ions, and the solid curves show the results
derived from the model. The insert shows the field dependence of the
strain amplitude $E_{13}^{}$ at 1.7 K derived from the experiments in
comparison with the calculated results.} \label{fig:xray}
\end{figure*}

The experiments were performed at HASYLAB using the BW5 high-energy
($\approx 100$ keV) beam-line with a triple-axis diffractometer and
$(1,1,1)$ Si/Ge-gradient crystals as monochromator and analyzer. The
single crystal of TmNi$_2$B$_2$C, grown by the flux-method
\cite{cho95}, is plate-like with irregular plate-shape and
approximate dimensions $4 \times 4 \times 0.8$ mm$^3$ along the $a$,
$b$ and $c$ axes, respectively. The crystal was mounted with the $a$
and $c$ axes in the scattering plane of a horizontal 10 tesla
cryo-magnet. Geometric constraints required that the field was offset
by $5^\circ$ from the $a$ axis in order to probe the expected lattice
distortion wave with the symmetry of the $\Delta_4^{}$-phonon mode
close to $(0.48,0,8)$ along with the $(0,0,8)$ Bragg reflection.
Temperature and field dependent superstructure peaks were found at
scattering vector $(0.484(2),0,8)$ with the integrated x-ray
intensities as shown in Fig.\ \ref{fig:xray}. In order to include the
intensity from minor satellite crystallites we use numerical
summation. The x-ray cross-section is completely dominated by the
scattering from the Tm ions, and scans performed at $(0.484,0,0)$
showed negligible intensity. This is clear evidence for a transverse
displacement of the Tm-ions along the $c$ axis corresponding to the
acoustic $\Delta_4^{}$-phonon symmetry. The $(0,0,8)$ Bragg
reflection was used as reference for accurate determination of the
superstructure peak position and estimates of the superstructure
displacement amplitudes.  However, the irregular crystal geometry and
the small scattering angles result in rather low and uncertain sample
transmission coefficients of the order of $4.6 \times 10^{-3}$ and
$8.0 \times 10^{-3}$ for the $(0,0,8)$ and $(0.484,0,8)$ reflections,
respectively. Including these corrections and defining the strain of
the Tm ion at site $\bm{R}_i^{}$
\begin{equation}
E_{13}^{}(i)=\frac{u_3^{}(\bm{R}_i^{})}{a/2}=E_{13}^{}
\cos(\bm{Q}_A^{}\cdot\bm{R}_i^{}+\phi)\,, \label{eq1}
\end{equation}
we find that the absolute displacements of the Tm ions at 1.7 K and
60 kOe, correspond to a strain amplitude $E_{13}^{}\approx 6 \times
10^{-3}$. The resolution of the diffractometer was determined by the
peak profiles of the $(0,0,8)$ Bragg peak yielding a crystal
mosaicity of $\approx 0.02^\circ$ FWHM. The scans along the $a$ axis
were essentially rocking curves with Lorentzian profiles while scans
along $c$ were Gaussian shaped. At low temperatures the correlation
length along $a$, corrected for instrumental resolution, is
determined to be 185 {\AA} when the quadrupolar phase co-exists with
the magnetic $\bm{Q}_F^{}$ phase, and it attains an almost constant
value of $\approx$ 275 {\AA} in the $\bm{Q}_A^{}$ phase. At zero
field and low temperatures the Gaussian peaks in the $c$ direction
indicate box-like structures of lengths $\approx$ 230 {\AA} that
increases to values above the resolution limit of $\approx$ 300 {\AA}
at 100 kOe. The superstructure peak widths are essentially
temperature independent at low temperatures but start to broaden when
approaching $T_Q^{}$.

Setting up a model we use the crystal-field (CF) parameters of the Tm
ions determined from the observation of CF transitions
\cite{gasser96} and from the high-temperature susceptibilities
\cite{cho95}. Here we take into account the anisotropy due to the
classical dipole--dipole interaction,
$\mathcal{J}_{aa}^{}(\bm{0})=\mathcal{J}_{cc}^{}(\bm{0})+9.15~\mu$eV,
adjust the CF-parameters accordingly ($B_2^0=-0.1$, $B_4^0=0.00033$,
$B_4^4=-0.01$, $B_6^0=0.78\times10^{-5}$ and
$B_6^4=-1.1\times10^{-4}$ in units of meV) and include the
demagnetization field in the Zeeman term. The zero-field value of
$T_{\text{N}}^{}$ determines the effective two-ion coupling to be
$\mathcal{J}_{cc}^{}(\bm{Q}_F^{}) \simeq8.7~\mu$eV, which is also
approximately its value at zero wave vector in the normal phase. The
purely magnetic part of the Hamiltonian is well characterized, and
the additional quadrupolar/magnetoelastic part of the mean-field
Hamiltonian is, by symmetry, considered to be
\begin{eqnarray}
\Delta\mathcal{H}(i)=&\!\!-\!\!&B_{13}E_{13}^{}(i)
O_2^1(i)+c_E^{}E_{13}^2(i) +{\textstyle\frac{2}{3}}
B_E^{}E_{13}^4(i)\nonumber\\
&\!\!-\!\!&K(\bm{Q}_A)\langle O_2^1(i)\rangle\left[O_2^1(i)-
{\textstyle\frac{1}{2}}\langle O_2^1(i)\rangle\right]. \label{eq2}
\end{eqnarray}
Here $O_2^1(i)=\frac{1}{2}(J_x^{}J_z^{}+J_z^{}J_x^{})_i^{}$ with $x$,
$y$, and $z$ along $[100]$, $[010]$, and $[001]$ respectively.
$E_{13}^{}(i)$ is determined by minimizing the strain-dependent part
of the average free energy:
\begin{equation}
\Delta F=-B_{13}^{}E_{13}^{}\langle\langle
O_2^1\rangle\rangle+{\textstyle\frac{1}{2}}c_E^{}E_{13}^2+{\textstyle\frac{1}{4}}
B_E^{}E_{13}^4\,, \label{eq3}
\end{equation}
where $\langle\langle O_2^1\rangle\rangle$ is the thermal value
$\langle O_2^1(i)\rangle\cos(\bm{Q}_A^{}\cdot\bm{R}_i^{}+\phi)$
averaged over the lattice. $E_{13}^{}$ determines $E_{13}^{}(i)$
according to eq.\ (\ref{eq1}).  The transition temperature $T_Q^{}$
of the quadrupolar ordered phase is independent of the value of the
fourth-order elastic constant $B_E^{}$. It is determined by the
effective quadrupolar coupling
\begin{equation}
K_{\text{eff}}^{}(\bm{Q}_A)=K(\bm{Q}_A)+\frac{B_{13}^2}{2c_E^{}}
\label{eq4}
\end{equation}
being equal to the inverse of the non-interacting quadrupolar
susceptibility at the temperature $T_Q^{}$, or
$K_{\text{eff}}^{}(\bm{Q}_A) =1/\chi_{O_2^1}^{}(T_Q^{})=0.0187$ meV,
when assuming $T_Q^{}=13.5$ K at zero field. The properties of the CF
Hamiltonian imply that the quadrupolar susceptibility increases with
the $a$-axis field, and the predicted field dependence of $T_Q^{}$ is
in reasonable agreement with observations according to Fig.\
\ref{fig:xray}. In the zero-temperature limit $\chi_{O_2^1}^{}$
becomes even more sensitive to the field. Here its maximum value at a
field of 20 kOe is nearly a factor of 3 larger than the zero-field
value. Although this increase is large, it is far from explaining the
dramatic enhancement of $E_{13}^{}$ observed at 1.7 K, see the insert
of Fig.\ \ref{fig:xray}. In the non-magnetic borocarbides, the
electron--phonon interaction is close to enforce a complete softening
of the $\Delta_4^{}$-phonon mode at the nesting wave vector
corresponding to $\bm{Q}_A^{}$ in the Tm system. The phonon energy is
10 meV at room temperature and about 4 meV at 4.2 K in the Lu
compound \cite{dervenagas95}. In the model above, the effects of the
electron--phonon interaction are included in terms of the
phenomenological parameters $c_E^{}$ and $B_E^{}$. If the
electron--phonon interaction is as important in the present system as
in Lu borocarbide, then $c_E^{}$ should be small, nearly zero, and
$B_E^{}$ would be decisive as soon as the lattice distortion becomes
non-zero. This is the argument for introducing $B_E^{}$, which has
the additional effect that the two contributions to the effective
quadrupolar coupling behave differently. This allows a
differentiation between the effects due to $B_{13}^{}$ and those due
to the quadrupolar--quadrupolar interaction $K(\bm{Q}_A)$.

\begin{figure}[b]
\includegraphics[width=0.99\linewidth]{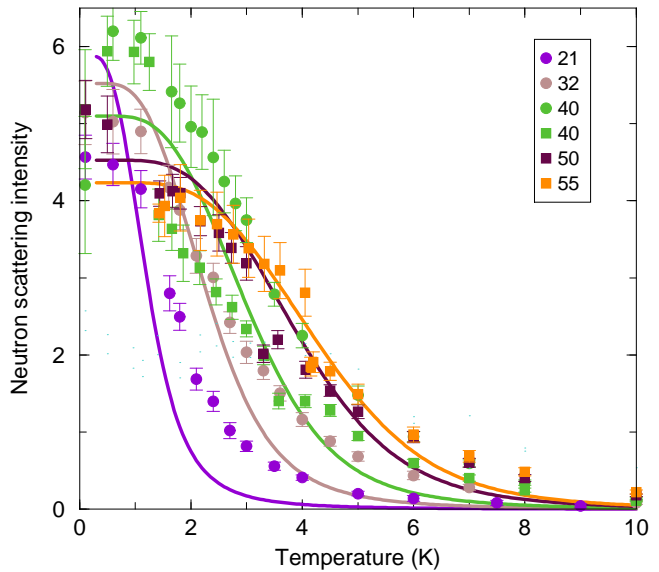}
\caption{The temperature dependence of the integrated neutron
scattering intensity at $\bm{Q}_A^{}=(0.484,0,0)$ at various values
of the applied field along $[100]$, see Ref.\ \cite{Katrine04}. The
different values of the field (in units of kOe) are specified  in the
box. The results were obtained in two different experimental set ups
(circles and squares). The repetition of the experiment at 40 kOe was
used for determining a common scale for the two sets. The intensity
is proportional to the squared amplitude of the transverse magnetic
moment at the wave vector $\bm{Q}_A^{}$ and the solid lines are the
calculated results.} \label{fig:Int}
\end{figure}

The quadrupolar ordering has the unique consequence (we are not aware
of any other systems showing this behavior) that the application of a
uniform field induces a modulated magnetic moment perpendicular to
the field. This effect was the very reason why we started to look for
the quadrupolar ordering, and the explanation for it is
straightforward. When $E_{13}^{}(i)$ or $\langle O_2^1(i)\rangle$ is
non-zero, the CF Hamiltonian includes an anisotropy term $O_2^1(i)$.
This implies that the paramagnetic moment of the $i$th site induced
by an $x$-axis field is not along the field, but is rotated an angle
towards the $z$ axis in proportion to the size of the anisotropy
term. Since the anisotropy, proportional to $\langle
O_2^1(i)\rangle$, is modulated in a sinusoidally way with the wave
vector $\bm{Q}_A^{}$, the $z$ component of the induced moment is
going to show the same modulation. The tetragonal symmetry of the
system implies that the Hamiltonian also includes the equivalent
terms obtained by interchanging $x$ and $y$ in eq.\ (\ref{eq2}). In
the domains in which $\langle O_2^{-1}(i)\rangle$ is ordered with the
wave vector $\bm{Q}_A'=(0,0.484,0)$, the application of a field along
[100] does not produce any perpendicular magnetic component, and the
quadrupolar order parameter is only weakly changed by the field. This
means that the [100] field induces a modulated $z$-axis moment at the
wave vector $\bm{Q}_A^{}$ but not at $\bm{Q}_A'$, and it destabilizes
the $\bm{Q}_A'$ domains. Hence, the first circumstance explains the
absence of magnetic scattering peaks at $\pm\bm{Q}_A'$ discussed in
Refs.\ \cite{Katrine00,Katrine04}. In the final comparison between
the model and the experiments, we have introduced one more parameter,
namely, the value of $\mathcal{J}_{cc}^{}(\bm{Q}_A^{})$. This
parameter is not decisive for the magnitude of $\langle
J_c^{}(\bm{Q}_A^{})\rangle$ (notice that $T_Q^{}\gg
T_{\text{N}}^{}$), but is important for $E_{13}^{}$ at the lowest
temperatures, and the fit is significantly improved, when assuming
this parameter to be small or, actually, slightly negative. The final
model is
\begin{eqnarray}
&&|B_{13}^{}|=1.91,~c_E^{}=360,~B_E^{}=4.8\times10^7\nonumber\\
&&K(\bm{Q}_A^{})=0.0086,~\mathcal{J}_{cc}^{}(\bm{Q}_A^{})=-0.0029\label{eq5}
\end{eqnarray}
all in units of meV. The different quantities have been calculated,
see Ref.\ \cite{JJ91}, by assuming a commensurable quadrupolar
ordering with a period of 25 lattice planes along the $a$ axis, i.e.\
$\bm{Q}_A^{}\equiv (0.48,0,0)$. The square of the distortions is
observed to be a factor of 2--3 smaller than calculated at fields
below 15 kOe. An equal population of the $\bm{Q}_A^{}$- and
$\bm{Q}_A'$-domains, at the lowest values of the field, may account
for a factor of 2. The model neglects any direct consequences of
superconductivity on the magnetic phase diagram, but the experimental
x-ray scattering intensities show kinks in their temperature
variations at the points where the field is equal to the
superconducting critical field $H_{c2}^{}$. This indicates a
(moderate) reduction of the quadrupolar order parameter in the
superconducting phase.

The neutron-diffraction results \cite{Katrine04} are compared with
theory in Fig.\ \ref{fig:Int} assuming that the value 6 on the
intensity scale corresponds to the moment amplitude of 5.0
$\mu_B^{}$. The calculated amplitudes of the modulation of the
moments are of the right order of magnitude. The maximum of the
amplitude in the zero-temperature limit is predicted to occur as soon
as the system leaves the $\bm{Q}_F^{}$-phase, at a calculated field
of 17 kOe to be compared with an experimental value of 10-15 kOe. The
experimental maximum is at about 40 kOe, but we have to remark that
there are relatively large, unexplained scatter in the experimental
results. Instead of focussing on the discrepancies we rather
emphasize the ability of the present model to account for the highly
disparate variation of the properties of the system. Using a simple
elastic model we estimate that $c_E^{}=360$ meV corresponds to an
acoustic phonon mode at 2.4 meV close to the soft phonon energy of 4
meV observed in the Lu compound. The value of $B_E^{}$ also appears
to be acceptable. The entropy of the electron--phonon system has
nearly vanished and the electron--phonon parameters $c_E^{}$ and
$B_E^{}$ are expected, as assumed, to stay practically constant below
20 K. Because of the small non-magnetic entropy, the heat capacity
anomaly at $T_Q^{}$ is expected to be weak, since the CF contribution
of the $4f$ electrons is negligible (the ground-state doublet remains
degenerate at zero field).

The analysis indicates that the electron--phonon system in
TmNi$_2$B$_2$C, like in the Lu compound, is very close to the point
of a lattice instability. In TmNi$_2$B$_2$C, the transition is
occurring because of the additional contributions of magnetoelastic
and quadrupolar--quadrupolar interactions. The magnetoelastic
interaction derives from the electrical crystal field produced by the
displacements of the ions, whereas the quadrupolar--quadrupolar
interaction may be mediated by the same charge-density wave
responsible for the strong electron--phonon interaction, and, hence,
may be an additional consequence of the Fermi-surface nesting. One of
the surprises of the present analysis is that the RKKY-exchange
interaction does not seem to be much influenced by the nesting, since
the interaction is found to be much smaller at the nesting wave
vector than at $\bm{Q}_F^{}$.

In conclusion, the x-ray experiments exposed the presence of a
quadrupolar phase in TmNi$_2$B$_2$C below 13.5 K at zero field. The
transition temperature increases and the accompanying deformation of
the lattice is grossly enhanced when a field is applied along
$[100]$. From the model analysis we conclude that this occurs mainly
because the lattice is already close to the critical point due to the
non-magnetic electron--phonon interaction. The magnetic interactions
are the final ingredients that enforce the system to enter the
quadrupolar phase, and the analysis indicates that the magnetoelastic
crystal-field and the quadrupolar--quadrupolar interactions are of
nearly equal importance,
$K(\bm{Q}_A^{})=0.46K_{\text{eff}}^{}(\bm{Q}_A^{})$. The quadrupolar
ordering has the unique effect that a field applied along $[100]$
induces antiferromagnetic ordering at the same wave vector as the
quadrupolar ordering. The Fermi-surface nesting is presumably
important for the strong electron--phonon interaction at
$\bm{Q}_A^{}$, and possibly also for the quadrupolar--quadrupolar
interaction. Contrary to this, the RKKY interaction is concluded to
be relatively weak at this wave vector. This suggests that those
electrons, which are close to the nesting areas and important for
creating the superconducting phase, are not the same as those
determining the RKKY interaction. The present discovery has basic
consequences for the understanding of the magnetism of the
borocarbides. It raises a number of questions: Is there a
strain/charge quadrupolar phase in the non-magnetic borocarbides or a
magnetic quadrupolar phase in other of the magnetic ones? Is the
strong peak at $(0.55,0,0)$ in the exchange interaction of the Er
compound, see Refs.\ \cite{Anette04,JJ02}, a consequence of a
quadrupolar interaction rather than the RKKY? Is the quadrupolar
interaction the reason for the longitudinal polarization
\cite{dervenagas96} of the antiferromagnetic ordered moments in the
Tb compound? {\it etc}.

This work is supported by the Danish Technical Research Council
via the Framework Programme on Superconductivity, and the Danish
Natural Science Council via Dansync and DanScatt. P.C.C. is
supported by the Director of Energy Research, Office of Basic
Energy Science under contract W-7405-Eng.-82.

\end{document}